\documentclass[conference]{IEEEtran}
\IEEEoverridecommandlockouts
\usepackage{cite}
\usepackage{amsmath,amssymb,amsfonts}
\usepackage{algorithmic}
\usepackage{graphicx}
\usepackage{textcomp}
\usepackage{xcolor}

\usepackage{tcolorbox}
\usepackage{booktabs}
\usepackage{enumitem}
\usepackage[breaklinks=true]{hyperref}
\usepackage{breakurl}

\newcommand{\RQone}{What do the modded markets and apps ecosystem look like, and what is its size?} %
\newcommand{\RQtwo}{What are the financial incentives for operating modded app markets? How does this affect the original developers and markets?} %
\newcommand{\RQthree}{What are the security implications of installing apps from these markets?}

\def\BibTeX{{\rm B\kern-.05em{\sc i\kern-.025em b}\kern-.08em
    T\kern-.1667em\lower.7ex\hbox{E}\kern-.125emX}}
\begin{document}

\title{ModZoo: A Large-Scale Study of Modded Android Apps and their Markets}

\author{\IEEEauthorblockN{1\textsuperscript{st} Luis A. Saavedra}
\IEEEauthorblockA{\textit{Computer Laboratory} \\
\textit{University of Cambridge}\\
Cambridge, United Kingdom \\
luis.saavedra@cl.cam.ac.uk}
\and
\IEEEauthorblockN{2\textsuperscript{nd} Hridoy S. Dutta}
\IEEEauthorblockA{\textit{Computer Laboratory} \\
\textit{University of Cambridge}\\
Cambridge, United Kingdom \\
hridoy.dutta@cl.cam.ac.uk}
\and
\IEEEauthorblockN{3\textsuperscript{rd} Alastair R. Beresford}
\IEEEauthorblockA{\textit{Computer Laboratory} \\
\textit{University of Cambridge}\\
Cambridge, United Kingdom \\
alastair.beresford@cl.cam.ac.uk}
\and
\IEEEauthorblockN{4\textsuperscript{th} Alice Hutchings}
\IEEEauthorblockA{\textit{Computer Laboratory} \\
\textit{University of Cambridge}\\
Cambridge, United Kingdom \\
alice.hutchings@cl.cam.ac.uk}
}

\maketitle

\begin{abstract}
We present the results of the first large-scale study into Android markets that offer modified or \emph{modded} apps: apps whose features and functionality have been altered by a third-party.
We analyse over 146k (thousand) apps obtained from 13 of the most popular modded app markets.
Around 90$\%$ of them are altered in some way when compared to the official counterparts on Google Play. Modifications include games cheats, such as infinite coins or lives; mainstream apps with premium features provided for free; and apps with modified advertising identifiers or excluded ads.
We find the original app developers lose significant potential revenue due to: the provision of paid for apps for free (around 5\% of the apps across all markets); the free availability of premium features that require payment in the official app; and modified advertising identifiers.
While some modded apps have all trackers and ads removed (3\%), in general, the installation of these apps is significantly more risky for the user than the official version: modded apps are ten times more likely to be marked as malicious and often request additional permissions.

\end{abstract}


Android has an open design philosophy, allowing users to easily install apps outside Google Play (sideload). 
Thus, alternative third-party markets have emerged that allow developers to share their apps in countries where Google Play is not present (China and North Korea), or does not allow paid apps and IAPs (In-App Purchases) (Cuba, Russia and Belarus).
There are open source markets such as F-Droid, and device manufacturers like Samsung, Huawei and Amazon may pre-install their own market app. 
\emph{Modded apps} are defined in this study as apps with code or metadata modified by an unauthorised developer or third-party. 
Therefore, \emph{modded markets} are app markets that 
focus on or advertise a large catalogue of predominantly modded apps.
Modded apps may (for free): 
unlock subscription features, provide infinite in-app or in-game currency, eliminate adverts and offer paid apps. 
This allows users to save money or try apps, games, and subscriptions before obtaining them from legitimate sources; to enjoy an ad-free experience; and to have an advantage over others or save time on games.

%
Modded markets are an important part of the Android ecosystem, offering millions of apps with clear benefits and desirable features to users. However, the extent of the modifications, security implications for users, and developer and market operator incentives are less obvious and so far unstudied. We fill this gap in knowledge by identifying 423 modded markets, studying their size, presence of ads and blogs, and ranking by popularity. We then analyse over 146k (thousand) apps and their metadata obtained from the 13 most popular modded markets over a 3-month monitoring period, and match these apps with their Google Play equivalents. This allows a direct comparison between modded and official versions.  
The larger modded app markets operate at scale, with an average of over 37k apps (max 221k) and around 2k apps added every fortnight.

They likely reduce developers' and official markets' income. Around 5\% of the apps are free copies of paid apps on Google Play, with a total value of USD \$33\,975 and estimated lifetime revenue in the Google Play (price $\times$ Google Play installs) of over \$2 billion.
Premium features usually charged via IAPs in popular apps are available for free in modded versions, e.g.\ ad-free audio in Spotify, which reports billions in IAP revenue per year~\cite{spotify_statista}. %
Also, 21\% of modded apps with ad IDs (advertiser IDs) have different ones to the official version in Google Play, and 6\% of modded apps include additional ad libraries, potentially redirecting ad revenue away from the original developer. 
Modded apps are riskier for users: many modded apps claim to remove ads but only 3\% do so, 23\% of modded apps request additional permissions, and nearly 9\% of apps are marked as malicious by VirusTotal, around 10 times the rate found in Google Play versions.
Sideloading is possible in consumer laptop and desktop operating systems and 
gives users more choice, allowing developers to sell apps and features without paying a percentage of revenue to official markets.
However, our work shows modded apps have significant negative effects. While third-party markets have the potential to benefit users and developers, some regulation is required to protect users and developer revenue streams.
This work is timely for regulators, who need to balance competition and fair markets with user and intellectual property protection.
The requirements to allow sideloading in the EU are being enforced and adapted by the EU's Committee on Internal Market and Consumer Protection (IMCO)~\cite{eu_dma,eu_dma_press}.
%
In summary, we make the following contributions:
\begin{itemize}[noitemsep,topsep=0pt]
   \item An overview of the modded app ecosystem and the first in-depth study of markets containing modded Android apps.
    \item Monitoring, data collection and analysis of 13 of the most popular modded markets over a three-month period, collecting 146k modded Android apps.
    \item We make our dataset, \emph{ModZoo}, available to other researchers.
    \item Matching modded apps with their Google Play counterparts, we find around 90\% of apps are modified in some form and 75\% have modified code. 
    \item The presence of these modded markets is likely to reduce income for app developers and official markets due to: the widespread availability of paid apps for free; premium features offered for free; and the redirection of ad revenue, including 21\% of apps with altered ad IDs.
    \item Modded apps are riskier for consumers: 23\% of modded apps requested additional permissions and nearly 9\% were marked as malicious by VirusTotal. %
\end{itemize}

\section{Research Questions and Methodology}

The main research questions this paper answers are:
\begin{itemize}[noitemsep,topsep=0pt]
    \item[RQ1] \RQone
    \item[RQ2] \RQtwo
    \item[RQ3] \RQthree
\end{itemize}

\subsection{Identifying and ranking modded markets}\label{sec:market_ranking}
We obtain a list of 423 sideloading and modded app markets by querying two popular search engines: Google Search and DuckDuckGo, 
using the following keywords in English, Chinese, Hindi and Russian: `Android app stores', `free Android app store', `mod \{apk/Android/games\}', `download premium apk', `download paid apps free', `\{paid/mod/premium\} apps for free', `unlocked android \{apps/games\}', `\{You\-Tube/\-Spotify/\-Trucaller\} mod'. 
Chinese and Russian were chosen due to the limited availability of Google Play in China and Russia. Hindi was added as preliminary results included Indian domains (`.in'). We manually verified the existence of apps advertised as modded apps in the markets. %

All 423 markets cannot be analysed in depth, thus a popularity-based ranking of the markets was curated using Google Trends. While only useful to compare the popularity of keywords over time, pair-wise comparisons for the 6-month period leading to our study allow us to obtain a relative ranking for all markets. We then cross referenced this ranking with the Tranco ranking corresponding to the 9-month period leading up to our study~\cite{tranco}. We found we analyse the top 7 most popular markets in the Tranco ranking, 9 out of the top 10, and other 4 markets within the top 35. Interestingly, out of the 423 markets, only 38 out of the top 60 in the Tranco list still offer modded apps three months later. The rest no longer operate or now focus on other activities such as offering news articles.

\subsection{Nomenclature}\label{sec:nomenclature}
The 146k \emph{modded apps} in our study each have a unique hash and correspond to 48\,384 unique package names, i.e. they are different modded versions of 48k unique apps.
We refer to \emph{exact matches} where we find an app one market with the exact same package name and version code as seen in another market.
Unless stated otherwise, we use exact matches for all our comparisons. 
We use the \emph{non-exact, latest-available match} when comparing a potentially malicious apps found on a modded market with the latest version of an app with the same package name on Google Play. 
Non-exact matches are a reasonable proxy when studying maliciousness as we assume later versions of the same app on Google Play are at worst similarly malicious to older versions. 
Non-exact latest-available matches are also the latest and only versions available in Google Play, so %
sections looking at app and IAP prices use the latest version metadata directly from Google Play as we were unable to find a reliable source of historic price data.
Modded APKs and their Google Play matches are analysed and the resulting profiles are stored for later comparison. We will refer to apps on Google Play which cost money as \emph{paid apps}, while any exact matches on modded markets are referred to as \emph{pirated apps} because they are offered without charge on modded markets.

In later sections we discuss five different types of app. \emph{Hash-identical} apps are those where the entire binary is hash-identical to their Google Play counterpart, i.e. where the entire packaged application (APK) is bit-for-bit identical, including manifest, libraries, code, etc. %
We also explore \emph{code-identical} apps: those whose code (.dex) files are the same, but other aspects, including permissions and manifest might differ. Similarly, \emph{certificate-identical}, \emph{permission-identical}, \emph{ad library-identical}, and \emph{ad ID-identical} apps, are those whose signing certificate, permission set, ad libraries set and advertising IDs are the same as found in their Google Play version, respectively.  
Their counterparts are \emph{code-modded}, \emph{certificate-modded}, \emph{permission-modded}, \emph{ad library-modded}, and \emph{ad ID-modded} apps. %

\subsection{ModZoo dataset collection}\label{android_scraping}
Our ModZoo dataset consists of 146\,162 downloaded modded apps, their metadata and analysis results as well as their 87\,792 exact and non-exact, latest-available matches from Google Play. 
We obtain Google Play apps from AndroZoo, a dataset %
which includes 21 million apps from Google Play, including different versions of the same app~\cite{androzoo}. %
We scraped the 13 most popular modded markets (see~\S\ref{sec:market_ranking}) between September and December of 2022 every 10-14 days to build our dataset of modded apps.
Our custom parallelised scrapers are written in Python3 to quickly obtain all relevant pages and APKs from the 13 modded markets. We used a set of proxies around the world to perform our data collection. 
Some of the scrapers use only HTML requests, while others also require Selenium and Mozilla's Gecko Driver to imitate user interaction. %
For other markets, we scraped their website first 
and then contacted the endpoints used by their custom market app. %
All information pages were stored, including the download pages, and all available modded APKs were downloaded.

We compute SHA256 hashes of all APKs to store each app with a particular hash only once. We map modded apps to their Google Play counterparts to enable a comparison between modded APKs and their official versions found in Google Play (AndroZoo).
ModZoo also includes the VirusTotal analysis results of 175\,584 APKs, including 103\,914 modded and 71\,670 Google Play APKs. The difference between the size of our ModZoo dataset and the number of VirusTotal analysis results is due to the use of existing results, as previous studies have found VirusTotal results to be more reliable after repeated scans~\cite{maat,rmvdroid}.

We make the ModZoo dataset available to the research community, dataset access can be gained at \url{https://www.cambridgecybercrime.uk/datasets.html}. 

\subsection{Static analysis methodology}\label{static_analysis_methodology_match_types}
Static analysis allows relatively quick results, ideal for the ModZoo dataset of more than 146k modded apps and their almost 88k Google Play counterparts. %

Our analysis pipeline starts by obtaining the latest data from AndroZoo. %
Then, it analyses the modded APKs in parallel, returning and storing their metadata and closest AndroZoo match, as well as whether it is an \emph{exact} or \emph{non-exact, latest-available} match (see~\S\ref{sec:nomenclature}). It then analyses the obtained AndroZoo match and stores the results.
We run the third-party reverse engineering tool Apktool~\cite{apktool} on each app and use the UNIX `keytool' command to obtain certificate information from each app. We obtain the certificate `owner', `issuer', `serial number', `certificate SHA256', `signature algorithm', etc.  
Running Apktool again creates the `apktool.yml' file, which we parse to obtain the APK's filename, minimum and target SDK versions, and version name and code. Our \textit{Manifest Parser} parses the `AndroidManifest.xml' file using a third-party Python XML library returning metadata attributes including the app's package name and version, permissions, activities, providers, receivers, intents, etc. 

To detect advertising libraries in the analysed APKs and their Manifest files, a `safelist' of ad library package names was created and iteratively extended as explained below. 
The \textit{Manifest Parser} analyses the `application', `meta-data' and `activity' attributes thoroughly, as this is where AppLovin and GoogleAds ad IDs, as well as the presence of IAPs can be found. %
We check whether the application attributes are present in our ad libraries safelist. If not in our list, it is added to a list of potential candidates to join the list, to be manually checked later. Thus, we have continuously expanded our safelist of ad libraries and reanalysed apps which analysis was older than the latest version of the safelist. All of the information gathered is stored as a profile in JSON format. The results returned to the analysis pipeline are: the package and version name, the JSON profile, ad IDs and ad libraries found, and ad library candidates. 
Then, the package name obtained from the manifest file and version code from the Apktool step are used to obtain from AndroZoo -- where available -- the \emph{exact} or \emph{non-exact, latest-available} match Google Play app.
The AndroZoo match app analysis follows the same steps, except the AndroZoo step is skipped. The modded app analysis results are stored, as well as those of their AndroZoo match.

\subsubsection{Modded apps Google Play matching}\label{matched_and_unmatched_apps}
A total of 136\,620 out of our 146\,162 downloaded modded apps were matched with a Google Play app present in AndroZoo using the methods described above. The 6.5\% unmatched modded apps correspond mostly to paid apps and games not available in AndroZoo. 
Out of those matched, 88.6\% are exact matches (same version number and package name), and only 11.4\% are non-exact, latest-available matches (same package name but the latest version number available at the time of the analysis).

\subsection{Ethics}\label{sec:ethical_considerations}
Our institutional ethics committee approved our study. 
Our ModZoo dataset contains publicly-available Android apps and their metadata, and is shared with other researchers after a thorough approval process. 
Apps were only collected for analysis, not used, except 
for the case study of 28 modded apps (see~\S\ref{sec:casestudyIAPS}). We only used the apps long enough to test the modded functionalities, used testing devices and accounts, using no personal data. 
We only installed the 28 apps one at a time through markets' websites or apps.
%
We did not undertake any activities which could affect other users, e.g. creating two accounts to look up the other user's account details in Truecaller. 
%
%
We contacted all market operators for comment using publicly-available contact details, stating our affiliation and purpose.

\section{The modded app ecosystem}
This section tackles RQ1: ``\RQone'' We leverage insights from our manual analysis of the 423 markets and the static analysis of our 146k app dataset obtained from the 13 most popular modded app markets. 

\subsection{Analysis of modded apps and markets}\label{sec:analysis_modded_apps_markets_further_analysis_appvn_androeed_5play}
Our technical analysis focused on the 13 highest-ranked modded markets, as determined in~\S\ref{sec:market_ranking}. Their average estimated size based on number of apps listed is 37\,486 apps with a mean of 15\,719 apps downloaded 
(see Table~\ref{tab:markets_big_table}). The difference is due to unavailability of some apps and broken download links. Markets marked with asterisks ($\ast$) in Table~\ref{tab:markets_big_table} were partially scraped as they label modded apps clearly.
While this made scraping them feasible, further analysis revealed apps not labelled as `modded', labelled `unmodded' or `original' are rarely hash-identical to their exact matches from Google Play, highlighting the inaccuracies of these labels. 
We obtained more than 10k app samples from `Appvn', `Androeed', and `5play' %
and compared them to those in AndroZoo based on their (SHA256) hashes since all signatures, metadata and code should be identical for unchanged apps. 
Out of the self-reported unmodded apps: %
Appvn had 35.8\% hash-identical apps (bit-for-bit identical to Google Play apps), higher than the 0\% found in the modded side of the market; Androeed had only 6.5\% hash-identical apps, up from 5.7\%; and 5Play 8.3\%, down from 10.0\% in the modded side of the markets.

\begin{table*}
\begin{center}
    \centering    
    \caption{Overview of the modded market ecosystem and proportion of modded apps.} 
    \label{tab:markets_big_table}
    \begin{tabular}{l r r r r r r r r r} %
        \toprule
         \multicolumn{1}{p{1.0cm}}{Market} & \multicolumn{1}{p{1.25cm}}{\raggedleft Estima\-ted Size} & \multicolumn{1}{p{1.2cm}}{\raggedleft Unique Apps} & \multicolumn{1}{p{1.3cm}}{\raggedleft Unique Packages} & \multicolumn{1}{p{1.0cm}}{\raggedleft Dupli\-cates} & \multicolumn{1}{p{0.8cm}}{\raggedleft Paid (\%)} & \multicolumn{1}{p{1.2cm}}{\raggedleft Modded Apps} & \multicolumn{1}{p{1.5cm}}{\raggedleft Unchanged Apps} & \multicolumn{1}{p{1.2cm}}{\raggedleft Modded Code} \\ %
         
         \midrule
         Appvn$\ast$ & 221\,039 & $\ast$4\,389 & $\ast$1\,866 & $\ast$8 & $\ast$5.0 & $\ast$4\,297 & $\ast$0 & $\ast$3\,586 \\ %
         RevDl & 42\,540 & 30\,477 & 9\,599 & 187 & 6.4 & 23\,217 & 3\,466 & 5\,068 \\ %
         HappyMod & 41\,385 & 26\,737 & 17\,249 & 12 & 3.7 & 19\,996 & 4\,098 & 12\,826 \\ %
         MODDROID & 34\,312 & 30\,738 & 17\,152 & 13 & 3.7 & 23\,316 & 4\,005 & 15\,153 \\ %
         APKMODY & 33\,914 & 10\,516 & 3\,081 & 214 & 4.5 & 8\,857 & 420 & 4\,550 \\ %
         androeed$\ast$ & 24\,252 & $\ast$15\,450 & $\ast$6\,869 & $\ast$6\,195 & $\ast$3.4 & $\ast$12\,069 & $\ast$731 & $\ast$9\,163 \\ %
         Rexdl & 22\,988 & 14\,262 & 5\,824 & 24 & 8.4 & 11\,666 & 1\,822 & 2\,621 \\ %
         5play$\ast$ & 19\,014 & $\ast$19\,674 & $\ast$15\,859 & $\ast$16\,203 & $\ast$8.1 & $\ast$16\,095 & $\ast$1\,610 & $\ast$9\,917 \\ %
         Malavida & 16\,519 & 19\,648 & 16\,128 & 16 & 0.0 & 14\,333 & 4\,115 & 3\,084 \\ %
         APKDONE & 11\,080 & 14\,908 & 3\,232 & 113 & 4.3 & 10\,099 & 139 & 7\,341 \\ %
         ApkVision & 8\,491 & 7\,983 & 6\,900 & 16 & 5.9 & 5\,632 & 1\,055 & 2\,683 \\ %
         LMHMOD & 7\,880 & 6\,865 & 4\,317 & 6\,303 & 3.7 & 5\,577 & 229 & 3\,597 \\ %
         An1 & 3\,906 & 2\,696 & 1\,198 & 44 & 4.0 & 2\,629 & 22 & 1\,661 \\ %
         \bottomrule
    \end{tabular}
   \end{center}
\end{table*}

Our smallest market analysed is `An1' with 2\,696 unique apps downloaded, and `Moddroid' is the biggest with $>$30k. Finally, `Appvn' has the biggest estimated size ($>$220k). 
The number of distinct apps (package names) is halved as markets often provide multiple versions of each app, unlike Google Play which only offers apps' latest version. 
Apps on these markets change frequently, with around 4k apps added weekly across all markets. However, 
around 25\% are hash-identical duplicates found in more than one market. 
`Duplicates' are apps advertised as different versions within a market which turn out to be hash-identical. 

The `Modded Apps' and `Unchanged Apps' columns present the number of apps 
that have been modified 
and those that are hash-identical copies of Google Play apps, respectively. %
Focusing on exact matches, we compute the number of code-iden\-ti\-cal and code-modded apps as defined in~\S\ref{android_scraping}: 81\,250 apps (68.1\%) have received changes to their code (`Modded Code' in Table~\ref{tab:markets_big_table}). %
Code-modded apps are closely related to permission-modded apps, as discussed later (see~\S\ref{sec:s3_permissions_adlibs_adids}).
Interestingly, although the markets focus on code-modded and modded apps, some of them have more code-identical than code-modded apps. This could be due to several reasons, the simplest being trying to offer a wider catalogue. 


\subsubsection{Categories and features}
\begin{figure*}
    \includegraphics[width=\textwidth]{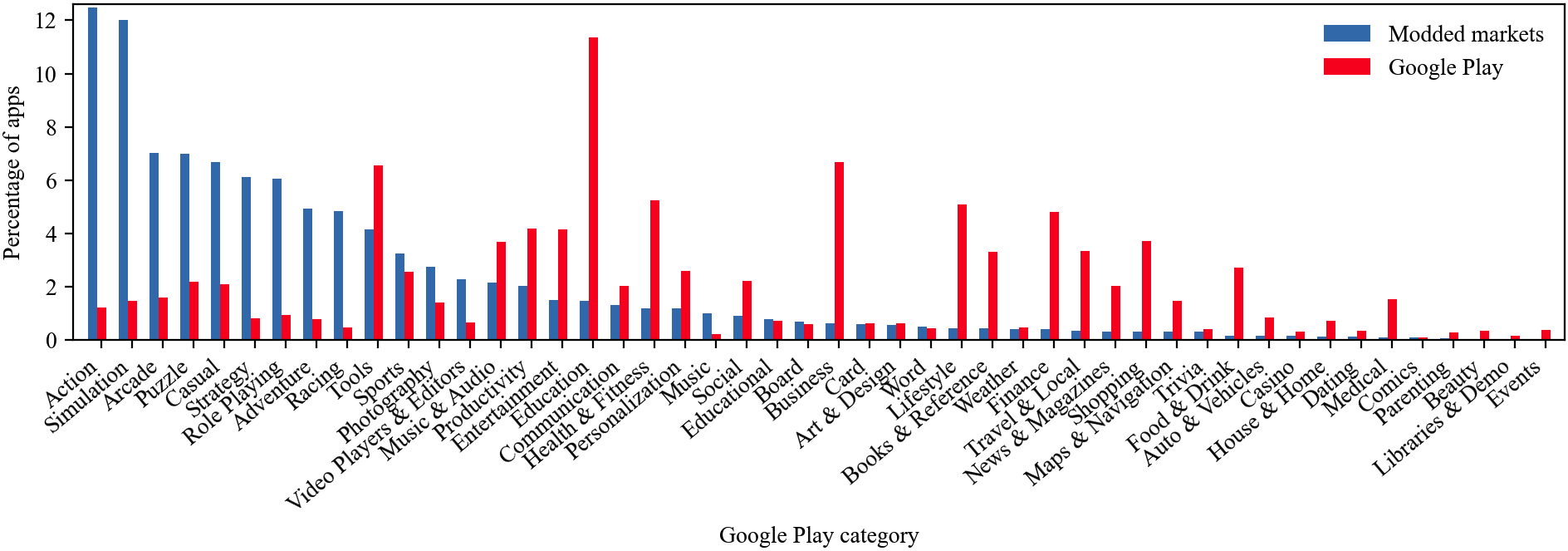}
    \caption{Distribution of Google Play app categories in the modded markets and Google Play}
    \label{fig:app_categories}
\end{figure*}

Modded markets focus heavily on games, we computed the Google Play categories of the modded app matches and those of a random sample of 100k Google Play apps.
As shown in Figure~\ref{fig:app_categories}, the 9 most popular modded app categories are game categories: `Action', `Simulation', `Arcade', `Puzzle', `Casual', `Strategy', `Role Playing', `Adventure' and `Racing'. 
Google Play categories, however, are led by `Education', `Business', `Tools', `Health and Fitness', `Lifestyle', `Finance', etc.\ most of which are at the tail end of modded app categories.

\subsubsection{Modded features}
Modded markets typically provide descriptions of modded app features to inform and entice potential users. %
The Android catalogue has gradually %
shifted towards `Freemium' apps~\cite{freemiumandroid}: apps with IAPs or subscriptions, and typically ads. Thus, the most popular modifications are associated with Freemium apps and games: mod money, 19\,206; unlimited money, 7\,202; free shopping, 3\,680; original, 3\,562; premium unlocked, 1\,257; full version, 1\,095; mod menu, 1\,020; mod premium, 895; mod unlocked, 730; unlimited coins, 702; no ads, 289.

\subsubsection{Paid (pirated) apps}\label{sec:paid_apps}
There are 6\,984 pirated apps in the 13 markets, corresponding to 2\,241 package names. %
The price distribution in Google Play is shown in Fig.~\ref{fig:paid_apps_and_iaps_cdf}, showing around 40\% have prices over \$5 in Google Play. Roughly 48\% of them have $>$500k Google Play installs. %
Their total value is USD \$33\,975, and \$9\,674 when counting each app (package name) only once. Their approximated lifetime revenue in Google Play (US price $\times$ number of global installs in Google Play) 
is \$2.28 billion. 
All modded markets studied lack payment mechanisms, thus the paid apps in Table~\ref{tab:markets_big_table} are available for free, and 
likely pirated copies of paid Google Play apps. The mean percentage of paid apps available for free across all markets is 4.7\%, with only `Malavida' hosting 0.0\% (6 apps in total).%
We estimate modded market operators would have spent at least \$9\,674 to get these paid apps from Google Play before hosting them in their markets if they had worked together and shared all their apps, or \$26\,901 if they had to buy each of the paid apps they host once.
It is possible market operators downloaded paid apps and requested refunds after making copies~\cite{googleplay_refunds}, resulting in \$0 of revenue for the original developers.

\subsubsection{In-App purchases (IAPs)}\label{sec:IAPS}
\begin{figure}%
       \includegraphics[width=0.47\textwidth]{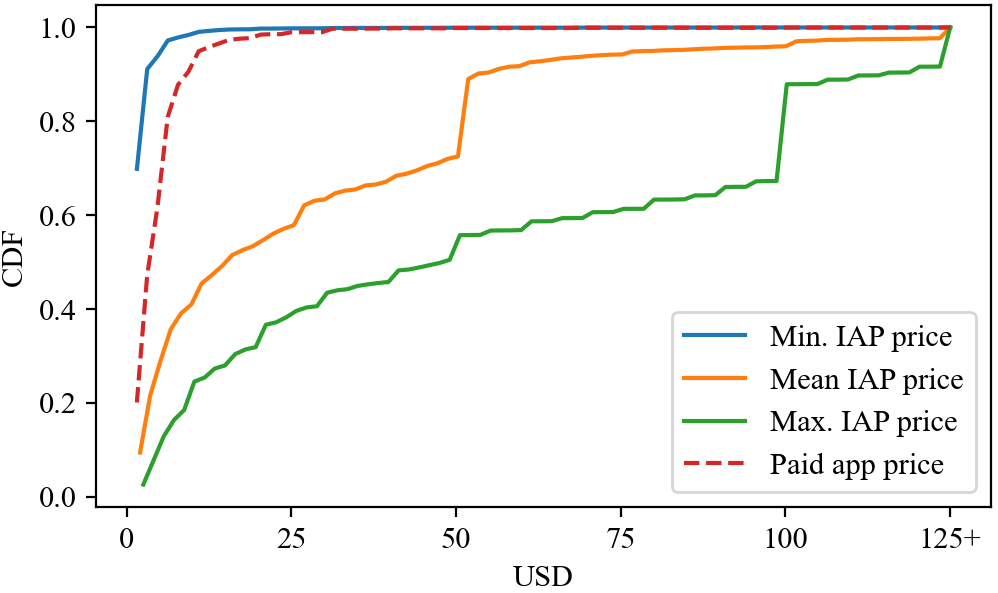}
    \caption{Google Play paid apps and IAPs price CDF.} %
    \label{fig:paid_apps_and_iaps_cdf}
\end{figure}

The 13 modded markets contain 100\,118 apps with IAPs worth at least \$3.7 million, with prices of up to \$1\,024 per item in Google Play. 
Over 40\% of the apps have IAPs with prices over \$100 in Google Play.
Their total price is at least \$3.7 million, based on the values reported on Google Play. %
Much IAP content and features are free in modded apps (see~\S\ref{sec:casestudyIAPS}).
The maximum, minimum and mean IAPs prices (in Google Play) are shown in Fig.~\ref{fig:paid_apps_and_iaps_cdf}.

\subsubsection{Countermeasures and market changes}%
During data collection we encountered different countermeasures employed by markets against scraping and automated downloads.
%
We observed that all markets analysed contain duplicate apps found in others. 
Waiting periods are common, preventing users from downloading multiple APKs in a short period, it provides an opportunity to show users more ads. Some markets used CAPTCHA tests, and a small number implemented Cloudflare DDoS and bot protection~\cite{cloudflare_2023}. 
Some markets introduced anti-scraping protections during our study. 
It is possible that our contacting markets for comment, scraping activity, or both made operators suspicious and more security-conscious. 
One market removed their 14 social media links (Github, LinkedIn, YouTube, etc.) from the English but not the Vietnamese version of their site during our study.%
Some markets require installing a proprietary app to download APKs they host, e.g.\ Moddroid, Jojoy, and HappyMod, which host mostly the same APKs in a shared back-end. 
Our scrapers obtain the metadata from the websites and contact the shared endpoints to download the APKs directly. 

\subsection{Case study}\label{sec:casestudyIAPS}
The five all-time most popular apps and games (as of March 2023) from Google Play are presented as a case study.
We manually test the official alongside modded versions from different markets to analyse their modded features and assess the scale of revenue loss caused by modded apps.
Google Play Protect is supposed to warn users of harmful apps on their devices, even when sideloaded. It may also deactivate or remove harmful apps. During our case study it warned ``Unsafe app blocked'' for 2 out of the 30 modded apps (28 plus 2 market apps), these were a game and the `Apkmody' market app. Users can click ``Install anyway''. 

Many modded apps showed a small logo or pop-up with the market's and sometimes modder's name. 
In some cases the market name displayed differed from the market we obtained the app from.
We found 14 out of the 28 app pairs studied had Google Mobile Ads and/or AppLovin ad IDs present. Of these, one TikTok and one Truecaller version had their Google Mobile Ads IDs removed. 

\subsubsection{TikTok} reported~\$1.5 billion IAP revenue in 2022~\cite{forbes_tiktok_iaps}, these are coins users can send to %
creators during livestreams, 
resulting in revenue for creators. 
They cost USD \$0.07--249.00 for 5--17\,500 coins. Modded versions claimed to offer unlimited coins, downloads without watermarks and no geolocation restrictions. Downloads worked well but coins were not included in any versions we tried. %

\subsubsection{SHAREit} premium features for \$1.99/month include no ads, exclusive customer service, regular cleanup and antivirus. Modded versions claim to remove all ads and include all premium features. None of the apps tested was ad-free, one provided regular cleanups and none provided premium customer service.

\subsubsection{Telegram} Premium for \$4.99/month, or \$35.99/year offers no ads and doubled limits (channel size, download speeds, document size, etc). Modded versions claim to provide these. Ads are only shown in public channels and were not served to us in genuine nor modded versions. Modded apps tested did not provide any other features, with download speeds as limited as free versions.

\subsubsection{Spotify} Premium for \$9.99/month offers no ads, higher sound quality, playing songs in any order, unlimited skips, downloads and offline listening. Spotify reported 2 million users ran modded versions in 2017 to avoid audio ads and subscriptions~\cite{spotifymoddedusers}. They reported a €11.57 billion revenue from Premium subscriptions and €1.7 billion ad revenue from non-premium users in 2023~\cite{spotify_statista}.
Modded versions advertise having the premium features. They were ad-free, provided unlimited skips and the ability to play any song. Not all make it clear they cannot provide downloads, offline listening and high quality audio. 

\subsubsection{Truecaller} Premium for \$4.99/month or \$49.99/year offers no ads, advanced spam blocking, seeing who viewed your profile, incognito mode, etc. Modded versions claim to have all premium features. However, although ad-free, the modded versions tested show all users as Gold members, with no effect for genuine users. Only one version showed who viewed or searched the user's profile. 

\subsubsection{Subway Surfers} `coins' and `keys' bundles cost \$0.99--99.99. Modded versions claim to have all these IAPs unlocked, some offer `God mode' game-play advantages: unlimited jumps, flying, etc. Piracy cost this game \$91 million by 2017~\cite{forbes_tapcore_piracy}. The versions we tested provided free IAPs and unlimited coins.

\subsubsection{Candy Crush Saga} offers many perks from \$0.99--99.99. 
All modded markets advertise having all levels unlocked, infinite lives, boosters, etc. Such offerings render all IAPs useless. The modded versions tested worked as advertised.

\subsubsection{Free Fire} `diamonds' cost \$0.99--49.99, perks \$8--12.99/month. Modded markets advertise diamonds and gameplay-related mods: aim-assist, no recoil, etc. Most versions tested did not work and the rest had none of these features.%

\subsubsection{My Talking Tom} `diamonds' cost \$1.99--99.99, perks \$4.99/month. Purchases remove all ads. Modded apps offer unlimited coins, perks and no ads. Unlimited coins unlock most content, but some ads and locked perks remain.

\subsubsection{Hill Climb Racing} perks and coins cost \$1.99--59.99, some remove ads. 
Modded versions advertise unlimited coins or all content unlocked. All versions tested provide unlimited coins, unlocking all content, although ads remained.

\subsection{Code, permissions and ad libraries}\label{sec:s3_permissions_adlibs_adids}
\begin{figure*}
    \includegraphics[width=\textwidth]{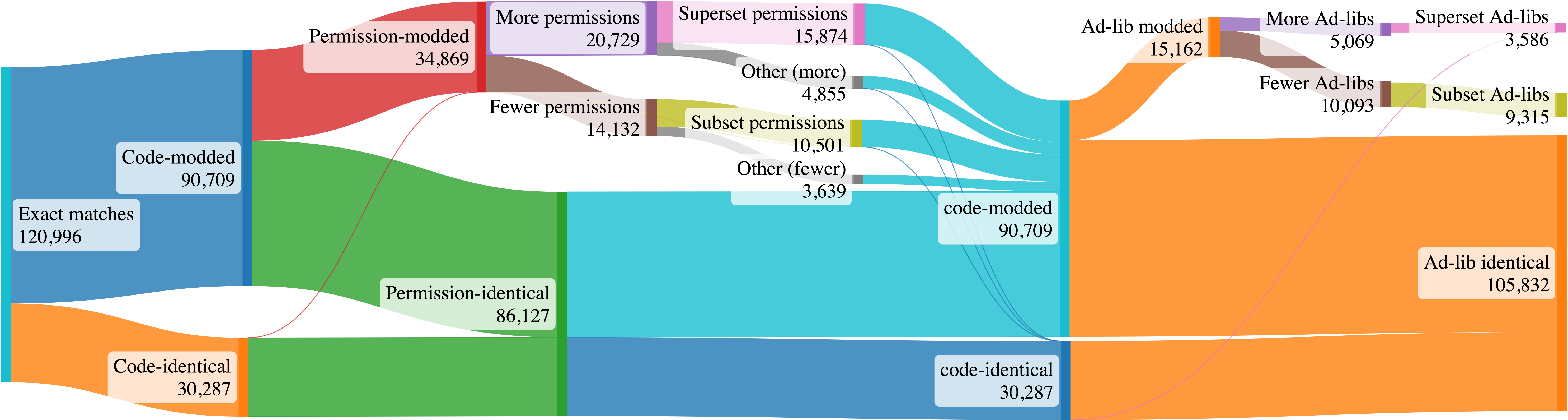}
    \caption{Permissions and ad library changes in code-identical and code-modded apps.}
    \label{fig:sankey_code_perms_adlibs}
\end{figure*}

We study changes to the code, stored in the `classes[n].dex' file(s) in relation to changed permission sets, ad libraries, and ad IDs for all exact matches. Of these pairs, %
a majority (75.0\%) are code-modded, 38.4\% of them are also permission-modded, with the majority (59.4\%) including additional permissions (as shown in Fig.~\ref{fig:sankey_code_perms_adlibs}).  
Some code-modded apps require further permissions for reasons related to the modifications, but the reason behind many additions was unclear.
Code-modded apps are mainly ad library-identical (83.3\%), only 11.1\% of them have fewer. 
In terms of ad IDs, 36.2\% of code-modded apps had none, and of those with ad IDs, 21.6\% had them changed. 
Altered ad IDs occur in a significant proportion of code-modded apps, we hypothesise permissions are sometimes added to code-modded apps in order to increase ad revenue for the modder.
For code-identical pairs, permissions and ad libraries remained completely unchanged in 99.9\%, and 100\% of the pairs, respectively. 
Their ad IDs either remained unchanged (65.3\%) or no ad ID was found. 
This suggests many apps are copied directly from Google Play without alteration, 
maybe to expand the catalogue and engage users even if a mod is unavailable. 
Many markets offer modded and original versions of each app, although we found inaccuracies in these labels (see~\S\ref{sec:analysis_modded_apps_markets_further_analysis_appvn_androeed_5play}).
Unmodified apps may be useful to users for compatibility or features.

\subsection{App signing certificates}
Most markets use mainly debugging and default Android Studio certificates unfit for app publishing. 
Some market-specific signatures such as `5play', which signs most code-modded apps with it. So does Apkmody, which includes the operator's name `Anh Pham' but mainly uses default signatures. Others use mainly `A1 Lazyland RU', present in most markets. Appvn uses mainly 5play.ru, all markets have 5play.ru and/or Apkmody certificates except Malavida.
All markets have a small proportion of third-party markets' and modders' certificates which include websites and Telegram links. Some were signed with `AntiLVL'. This analysis confirmed our previous findings of cross-market duplicate apps. 

\begin{tcolorbox}[left=1mm, right=1mm, top=1mm, bottom=1mm]
This section looked at pirated apps, 
IAPs and a case study of popular apps and games. %
Most popular categories and modded features relate to games. 
Many apps are code-modded, bypassing subscription features or in some cases removing ads. Others added permissions, ad libraries, and changed ad IDs. `Unmodded' labels and `ad-free' app descriptions cannot be trusted. 
\end{tcolorbox}

\section{Market operator motivations and income}
This section tackles RQ2: ``\RQtwo'' We approach these based on our observations and analysis results to analyse possible revenue streams in modded markets and operators' economic incentives. 

\subsection{Blogs, sponsored posts and ads}\label{blogs_markets_pricing}
We manually studied blogs, sponsored posts and ads in the 423 modded markets. Blogs are present in a third of them. They host articles about modded app updates, installation guides, and often also news articles, tips and tricks, rankings, and product or app reviews. Many markets openly displayed their pricing for advertising through different ads, product reviews, or guest posts. Others were open to contact and 
only a minority did not accept sponsored posts or ads.
Some blogs are inactive and 13\% have 5 or fewer posts. 
One market priced sponsored posts and ads at USD \$250--300; others for \$100. 
Most had lower prices starting at around \$30 for general posts, \$45 for casino-related posts, and more for those related to ``gambling, adult, dating, vaping, CBD, or cannabis''. %
Most posts are admin-uploaded, making it difficult to count the sponsored posts.
Some offered different ad types including sidebar and pop-ups for \$50--200/month. %

\subsection{Advertising libraries and advertiser IDs}
%
Many code-modded apps are advertised as ad-free versions of `Freemium' apps. We use a safelist (see~\S\ref{static_analysis_methodology_match_types}) to confirm whether they are ad-free and what other changes they have. %
Google Mobile Ads and AppLovin ad libraries (two of the most popular) include their ad ID in the manifest file, allowing us to compare them between modded and original apps.
We found 20.5\% of modded apps with ad IDs had them altered. 
It appears to be widespread practice to redirect ad revenue from the original developer to the modders or modded markets. 
Also, 41\,321 apps use ad libraries other than Google Mobile Ads and AppLovin, and in total 10\,990 apps have changed ad libraries compared to their Google Play version. %

The most popular advertising and tracker libraries present in our modded apps are GoogleAds, Facebook, and Unity, %
followed by AppLovin, ironSource, Vungle, AdColony, Tapjoy and InMobi. %
Their relative popularity is mostly the same in modded apps and their Google Play counterparts. Providers most affected by `ad-free' modded apps are GoogleAds (20.3\%), Facebook (14.9\%), Unity3D (9.6\%), and AppLovin (8.1\%). %
We found 
10\,353 contained no ad libraries originally, 4\,180 (2.86\%) had all removed and 2\,636 had their AppLovin and GoogleAds ad IDs removed. 
So, while some modded apps have had ad libraries removed, they are in the minority. %
The presence of libraries implies the possibility of ads in an app, e.g.\ the popular Unity library used in many games (`com.unity3d') can be used to display ads. 
Thus, we may have underestimated the number of ad-free apps. 
Further dynamic, manual analysis would be needed to confirm this, which would be impractical given the scale of our dataset.

\subsection{Advertising libraries, advertiser IDs and permissions}\label{sec:permissions_adlibs_adids}
\begin{figure*}
    \includegraphics[width=\textwidth]{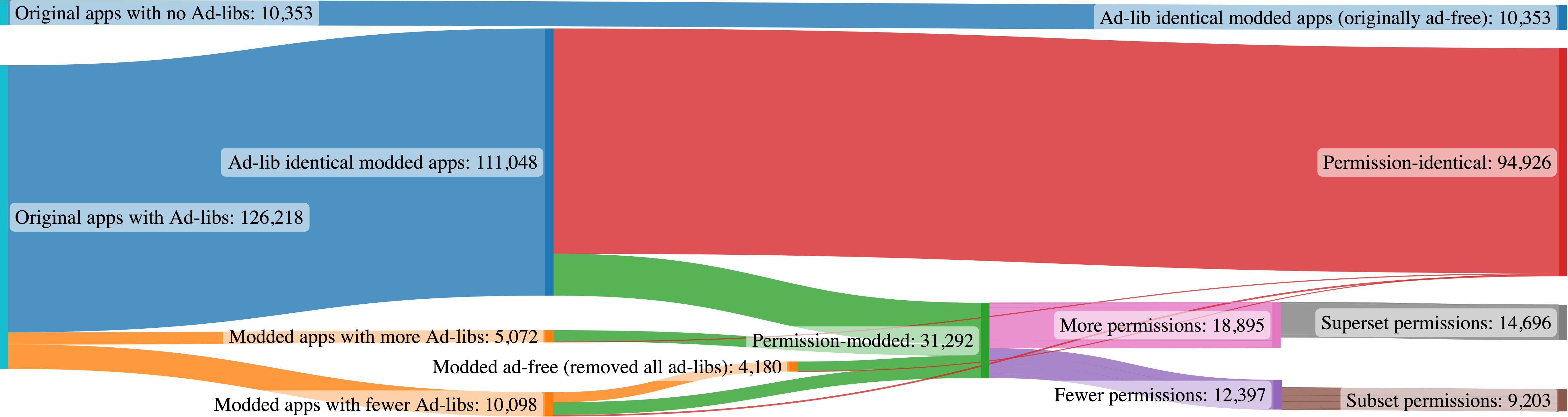}
    \caption{Distribution of ad libraries in original and modded apps and modded permissions.}
    \label{fig:sankey_adlibs_perms}
\end{figure*}

Changes to permissions have security implications. Combined with the aspects already presented in relation to ad libraries and ad IDs, they might provide increased revenue to modders or market operators. E.g.\ an ad library might use added location permissions to display more relevant ads.
The small proportion of ad-free apps (2.86\%), those which contain no ads where their Google Play counterparts do, typically present fewer permissions (88.9\%), a strict subset of the original permissions (76.2\%), and only 4.2\% are permission-identical, as shown in Fig.~\ref{fig:sankey_adlibs_perms}.
Thus, there is a genuine small offering of ad-free versions of popular apps with smaller permissions sets. 
Furthermore, as shown in Fig.~\ref{fig:sankey_adlibs_perms}, 91.0\% of modded apps are ad-library-identical to their Google Play counterparts. 
These tend to be permission-identical apps (84.3\%), with the rest mostly having more (11.4\%) and a superset (10.3\%) of permissions. Apps with added ad libraries are mainly permission-modded apps (91.3\%), with 64.9\% of them having more permissions. %
Those with removed ad libraries are mostly permission-modded (88.6\%), mainly with fewer permissions (60.9\%) or a strict subset (44.1\%). %
These results show ad libraries are not typically changed and are closely linked to permissions. 
\begin{figure*}
    \includegraphics[width=\textwidth]{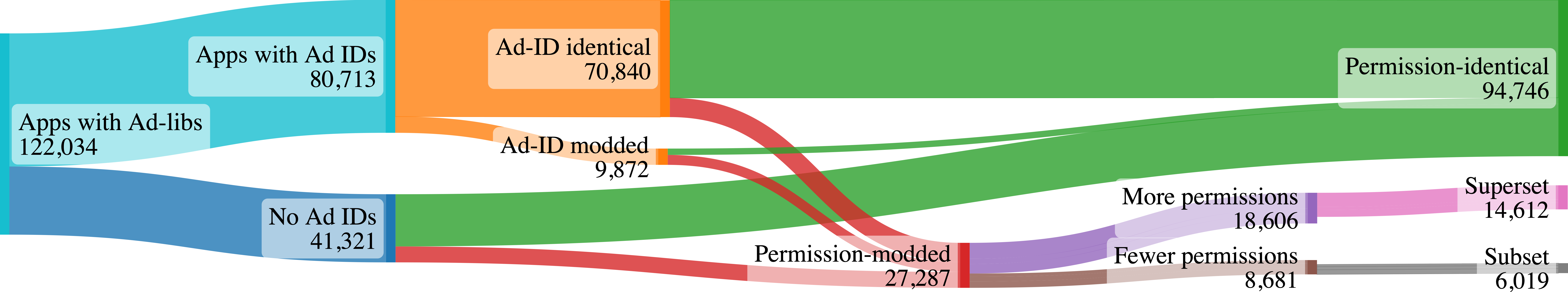}
    \caption{Distribution of permissions and advertiser IDs in modded apps with ad libraries.}
    \label{fig:sankey_ad-ids_perms}
\end{figure*}
When focusing on their GoogleAds and AppLovin ad IDs, 58.0\% of modded apps with ad libraries have unchanged ad IDs, 8.1\% had changed ad IDs, and 33.9\% of the modded apps had no ad IDs.
Ad-ID-identical apps were mostly permission-identical (83.6\%), with another 13.6\% having more permissions, as shown in Fig.~\ref{fig:sankey_ad-ids_perms}. Ad-ID-modded apps, however, were permission-modded in 61.1\% of the cases, with an even split of more and fewer permissions. %
Those with no ad IDs were mainly permission-identical (76.7\%). %
This suggests again a strong correlation between permission-modded, ad-library- and ad-ID-modded apps. This could be due to modders wanting to be compensated with ad revenue or more permissions giving more granular data to ad libraries, increasing ad revenue. %

\subsection{Modded apps and user displacement}
Piracy has been found to have a negative impact on revenue in the music and media industries~\cite{rob2006piracy,smith2012piracy}. 
Tapcore estimated 14 billion app installs were pirated in 2017, costing app developers more than \$17.5 billion, and Subway Surfers \$91 million~\cite{forbes_tapcore_piracy}. It is difficult to estimate the current revenue loss with the growth of IAPs and ad revenue in apps and games since 2017.
Studies on computer game piracy found high displacement rates of -2.49 for games, meaning each illegal download of a game typically displaces multiple genuine purchases~\cite{global_piracy_study}. Unlike for music, films, series, and books, where pirates tend to increase legal consumption while decreasing illegal consumption, 
they found game pirates increase or maintain illegal consumption over time, resulting in user displacement and lost revenue. 

%
\begin{tcolorbox}[left=1mm, right=1mm, top=1mm, bottom=1mm]
We identified possible revenue streams for market operators and modders including ads and sponsored posts. Our analysis shows a correlation between ad-library-modded and permission-modded apps. Code-modded apps are typically permission-modded and some ad-library-modded. Code-identical apps are permission and ad-library-identical. Furthermore, 1 in 5 code-modded apps with ad IDs had them changed. %
We found 6\,984 pirated apps with \$2.28 billion in estimated lifetime revenue in Google Play, and 100k apps with IAPs typically offered for free in modded apps (see~\S\ref{sec:paid_apps}--\S\ref{sec:casestudyIAPS}). Modded apps may cause user displacement, disrupting original developers' and markets' revenue and innovation. 
\end{tcolorbox}

\section{Security implications of modded apps and markets}
This section answers RQ3: ``\RQthree'' %
App analysis and VirusTotal malware analysis results are combined to tackle this from the consumers' perspective. However, there are also security and economic implications for the original app developers. Many of the modded apps use the original API keys so original developers pay for cloud services and API calls, etc. 
(see~\S\ref{sec:casestudyIAPS}). 
Also, other users' security might be affected. Modded apps can change what users can or cannot see in social networking apps, potentially exposing other users’ information beyond their preferences. 
It may also affect other Internet users, e.g. if modded apps embed a botnet. 

\subsection{VirusTotal analysis}
The VirusTotal analysis methodology is based on previous approaches, VirusTotal is queried with the hashes present in the entirety of ModZoo obtaining all existing analysis results. Analysis results obtained after repeated scans have been found by previous studies to be more reliable than new results~\cite{maat,rmvdroid}. 
Similarly, the recommended threshold of around 10\% of antivirus engines (AVs) flagging APKs as malicious is used (see~\S\ref{sec:related_work}). %
Furthermore, existing~\cite{avclass_paper} and custom tools were used to obtain unified malicious labels. %
%
We use VirusTotal to get insights into the entire ModZoo dataset. More advanced techniques could be used on a random or selected sample of the dataset, but that is considered future work.

Modded apps are sometimes paired with the non-exact, latest-available matches when the exact version of the app is not in AndroZoo. This is reasonable since the latest-available version on Google Play should be just as safe or safer than older versions. AndroZoo has most app versions, its authors have mitigations for robust scraping~\cite{androzoo}. 

\subsubsection{Malware, adware, and PUPs} %
\begin{figure*}
    \includegraphics[width=\textwidth]{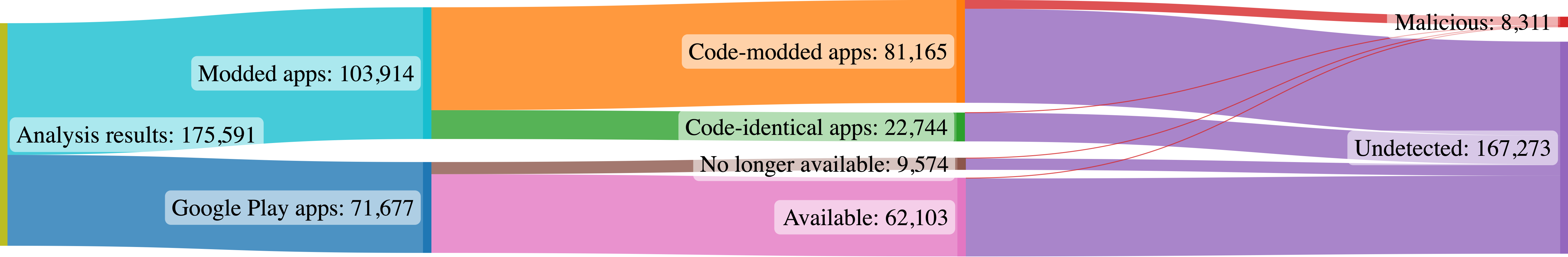}
    \caption{Distribution of malicious apps across Google Play and modded apps.}
    \label{fig:sankey_vt_results}
\end{figure*}

The VirusTotal results cover 103\,914 of the modded apps from our ModZoo dataset and 71\,670 Google Play (AndroZoo) apps. 
We found almost 9\% of code-modded apps and only 0.5\% code-identical apps coming from modded markets were labelled malicious compared to only 0.9\% of their currently-available Google Play counterparts, as shown in Figure~\ref{fig:sankey_vt_results}. 
Users are more vulnerable to malware, adware, potentially unwanted programs (PUPs) and other malicious programs when downloading modded apps. %

In total, 167\,273 apps were marked as undetected and 8\,311 (4.7\%) as malicious. Of these, 85.3\% %
came from modded markets and the rest from Google Play (AndroZoo). This translates as 6.82\% of modded apps and 1.70\% of Google Play apps in ModZoo classified as malicious. However, 8.59\% of code-modded apps are malicious, against only 0.51\% of code-identical apps.
The risk of modded apps goes beyond this, since many of the apps offered in modded markets are no longer offered in Google Play: 13.36\% of the Google Play counterparts are no longer available as of March 2023. Of these, 6.72\% are marked as malicious, compared to only 0.93\% of those still available. 
For hash-identical apps the risk is obviously identical between Google Play and modded markets while both host the app. However, Google analyses apps and responds quicker to incidents and user reports, providing better security protections than modded markets.

Malicious apps are flagged as PUPs such as Lucky\-Patcher, used to modify Android apps, and many are flagged with more worrying Trojan-like malware such as Andreed, Triada, RemoteCode, HiddenAds, Kyvu, (LuckyPatcher) IBGV, etc. and more general labels as `downloader', `virus'. 
The 20 most prominent labels are shown in Table~\ref{tab:virustotal_common_threat_labels}, 
4 of them are not present in Google Play apps at all, while most others have a significantly bigger presence in modded apps.

\begin{table}[htbp]
    \caption{Most common VirusTotal threat labels and their distribution}
    \label{tab:virustotal_common_threat_labels}
    \begin{tabular}{lrrr}
        \toprule
         Threat Label & Total & Modded Apps & Google Play \\ %
         \midrule
         None & 163\,788 & 93\,541 & 70\,247 \\
         andreed & 3\,830 & 3\,761 & 69 \\
         grayware & 3\,586 & 3\,052 & 534 \\
         fyben & 2\,111 & 2\,078 & 33 \\
         adware & 1\,011 & 490 & 521 \\
         downloader & 465 & 413 & 52 \\
         androeed & 183 & 181 & 2 \\
         kyvu & 72 & 19 & 53 \\
         grayware:tool & 51 & 37 & 14 \\
         triada & 28 & 7 & 21 \\
         remotecode & 27 & 24 & 10 \\
         dataeye & 24 & 14 & 10 \\
         hiddenads & 23 & 18 & 5 \\  
         luckypatcher & 21 & 21 & 0 \\
         ibgv & 18 & 18 & 0 \\
         spyware & 18 & 16 & 2 \\
         tencentprotect & 18 & 10 & 8 \\
         fleeceware & 17 & 8 & 9 \\
         boogr & 14 & 13 & 1 \\
         wamod & 14 & 14 & 0 \\
         virus & 13 & 13 & 0 \\
        \bottomrule
    \end{tabular}
\end{table}%

\subsubsection{Permissions in code-modded apps}
Many dangerous permissions are added to code-modded apps. Malicious code-modded apps have a higher incidence of these as shown in Table~\ref{tab:top_added_permissions_codemoddedmalicious_and_codemodded_apps}, with 14.6\% adding `SYSTEM\-\_ALERT\-\_WINDOW' which allows creating windows on top of any other app, which can be used for phishing attacks. A further 9.0\% malicious code-modded apps added the `READ\-\_EXTERNAL\-\_STORAGE' permission, which allows access to other apps' files in the MediaStore, potentially exposing users' personal data. Malicious code-modded apps are twice as likely to request these, although there might be genuine need for some of them in modded apps. The following permissions are more than 4 times more likely to be used in malicious than non-malicious code-modded apps: `WRITE\-\_SETTINGS' which allows apps to read system settings, `READ\-\_LOGS' which is not to be used by third-party apps since it allows apps to ``read the low-level system log files'', which may contain users' private information, and `CAMERA'. Other risky permissions such as `ACCESS\-\_COARSE\-\_LOCATION' and `ACCESS\-\_FINE\-\_LOCATION' are less common but are 8 times more likely to be added to malicious apps. Google distinguishes different permission protection level categories: `dangerous', `normal', and `signature'.~\footnote{\url{https://developer.android.com/reference/android/Manifest.permission}} 
It is worth noting that although `normal' permissions do not require user confirmation in-app like `dangerous' permissions, they are still potentially dangerous as users of modded markets are not presented with accurate information of the `normal' or `dangerous' permissions used by apps.
They are all android.permission.\{\} except `net.dinglisch.\-android.\-tasker.\-PERMISSION\_RUN\_TASKS' and `com.android.\-launcher.\-permission.\-INSTALL\_\-SHORTCUT'. 
We classified them as dangerous and normal, respectively, although they are not included in Google's classifications.


\begin{table*}[htbp]
    \caption{Top 30 added permissions in malicious code-modded and code-modded apps and their category.} 
    \label{tab:top_added_permissions_codemoddedmalicious_and_codemodded_apps}
    \begin{tabular}{llcc}
        \toprule
        Permission & Category & Malicious Code-Modded (\%) & Code-Modded (\%) \\ 
        \midrule
        android.permission.SYSTEM\_ALERT\_WINDOW & signature & 14.60 & 8.72 \\
        android.permission.READ\_EXTERNAL\_STORAGE & dangerous & 9.01 & 4.10 \\
        android.permission.BLUETOOTH\_ADMIN & normal & 7.39 & 1.65 \\
        android.permission.BLUETOOTH & normal & 7.19 & 1.62 \\
        android.permission.WRITE\_SETTINGS & signature & 7.05 & 1.70 \\
        android.permission.CHANGE\_WIFI\_STATE & normal & 6.68 & 1.55 \\
        android.permission.FLASHLIGHT & normal & 6.61 & 1.56 \\
        android.permission.USE\_FINGERPRINT & normal & 6.59 & 1.57\\
        android.permission.READ\_LOGS & very dangerous & 6.50 & 1.54 \\
        android.permission.REQUEST\_IGNORE\_BATTERY\_OPTIMIZATIONS & normal & 6.50 & 1.56 \\
        android.permission.READ\_SETTINGS & uncategorised & 6.50 & 1.56 \\
        net.dinglisch.android.tasker.PERMISSION\_RUN\_TASKS & dangerous & 6.50 &  1.56\\
        android.permission.CAMERA & dangerous & 6.38 & 1.54 \\
        android.permission.REQUEST\_INSTALL\_PACKAGES & signature & 5.62 & 1.07 \\
        android.permission.VIBRATE & normal & 4.38 & 0.88 \\
        android.permission.WRITE\_EXTERNAL\_STORAGE & dangerous & 3.82 & 3.01 \\
        android.permission.ACCESS\_WIFI\_STATE & normal & 3.41 & 0.86 \\
        android.permission.QUERY\_ALL\_PACKAGES & normal & 2.81 & 6.66 \\
        android.permission.GET\_TASKS & normal & 1.96 & 0.74 \\
        android.permission.READ\_PHONE\_STATE & dangerous & 1.38 & 0.63 \\
        android.permission.RESTART\_PACKAGES & deprecated & 1.06 & 0.13 \\ 
        android.permission.KILL\_BACKGROUND\_PROCESSES & normal & 1.01 & 0.11 \\
        android.permission.RECEIVE\_BOOT\_COMPLETED & normal & 0.90 & 0.15 \\
        android.permission.CHANGE\_NETWORK\_STATE & normal & 0.85 & 0.10 \\
        android.permission.BATTERY\_STATS & signature & 0.78 & 0.09 \\
        android.permission.ACCESS\_COARSE\_LOCATION & dangerous & 0.76 & 0.09 \\
        android.permission.BROADCAST\_STICKY & normal & 0.76 & 0.07 \\
        android.permission.ACCESS\_FINE\_LOCATION & dangerous & 0.67 & 0.09 \\
        com.android.launcher.INSTALL\_SHORTCUT & normal & 0.67 & 0.09 \\
        android.permission.DOWNLOAD\_WITHOUT\_NOTIFICATION & normal & 0.53 & 0.18 \\
        \bottomrule
    \end{tabular}
\end{table*}

\begin{tcolorbox}[left=1mm, right=1mm, top=1mm, bottom=1mm]
The security of modded markets is significantly lower than that of 
Google Play, with 8.6\% of code-modded apps and 6.8\% of apps overall flagged as malicious by VirusTotal, against %
0.9\% of currently-available Google Play apps. 
Furthermore, we found a high number of dangerous and risky permissions in code-modded apps, especially those classified as malicious. %
\end{tcolorbox}

\section{Limitations}\label{limitations_section}
%
The analysis of ad libraries in modded apps is potentially biased, as the accuracy of results depends on that of our safelist. %
Thus, we expanded it periodically as more apps were analysed.
Code obfuscation and shrinking are techniques available to developers to make apps more secure, difficult to reverse engineer, and storage efficient. However, they also undermine static analysis of ad libraries. Ad IDs, permissions, and other parameters studied are not affected by this limitation. Several analysis tools have been proposed to study the libraries present in obfuscated apps~\cite{libid,orlis}, 
They are infeasible considering the scale of ModZoo as they require downloading all libraries of interest. A safelist is an acceptable compromise between accuracy, speed and scale since code obfuscation and shrinking are not enabled by default in Android Studio. 
%
%
Google Play reports IAP prices as ranges instead of their number, price and which are subscriptions. %

\section{Related work}\label{sec:related_work}
Wang et al.\ analysed 6 million Android apps in 16 Chinese markets, inter-market similarity, publishing behaviours, malicious and fake apps~\cite{wang2018chinesemarkets}. The markets performed substantially worse than Google Play. We found similar in modded market security and presence of pirated apps. Also, we explored operator and modder motivations and revenue streams. Others studied Android app attribution, and found the lack of metadata in AndroZoo a limitation to study app attribution at scale~\cite{mixedsignals}. Metadata for removed Google Play apps is lost, thus we stored modded market app metadata in our ModZoo dataset. %

Other studies focused on Android VPN~\cite{ikram2016analysis} and firmware over-the-air~\cite{blazquez2021fota} apps, analysing their security, permissions and presence of malware through VirusTotal.

Others found free games in Google Play have 3.4 times more trackers and twice the number of dangerous permissions as paid ones~\cite{pricetoplay_laperdrix}. %
Kumar et al.\ analysed differences in 26 countries' Google Play markets, %
finding apps are often unavailable due to developer-introduced geoblocking~\cite{geodifferenceskumar}. %
These aspects could motivate modders and users.

Shen et al.\ found malicious apps last more than twice as long on Google Play than manufacturer-provided markets~\cite{malicious_apps}. We found the opposite for modded markets, as operators lack the motivations device manufacturers have to keep their platforms secure. Our study is also novel in the mapping of third-party (modded apps) with their Google Play counterparts to compare ad libraries, permissions sets 
and security implications. Others found repackaged apps aimed at tricking users to think they are genuine apps are common in official markets and half of them contained adware~\cite{repackagedapps2019}. Instead, modded apps are advertised as modified versions. 
They found half of the 15k repackaged apps contained adware, against our 9\% malicious code-modded apps. However, only 4\% of them added permissions against 24\% of code-modded apps in our study. They did not study ad IDs and their results are not reproducible due to their dataset's unavailability.
%

Previous research separated prominent and trivial permissions~\cite{aswini2014droid}, 
created permissions graphs to find outliers~\cite{sokolova2017android,taha2021hybrid}, and 
found malware-related permissions based on other datasets~\cite{alswaina2018android}. They rely on existing datasets or do not share their own. Unlike ours, they do not consider the connections between ad libraries and permissions changes. We also explored permissions added to malicious code-modded apps and found increased use of dangerous permissions.
Others studied manifest file intents and context~\cite{li2016android,suarez2017droidsieve}, while some identified packages and APIs used~\cite{aafer2013droidapiminer}. Static analysis is common to these large-scale approaches. We combine it with market and VirusTotal analysis. %

Previous studies have used VirusTotal to analyse apps at scale. 
Zhu et al. surveyed 115 papers to identify VirusTotal analysis methodologies~\cite{10.1145/3372297.3420013}. They collected executables analysis results for a year and found the threshold approach (labelling files malicious when flagged by at least $N$ antivirus engines included in VirusTotal) outperforms others. Most papers use thresholds to classify malicious files and the most popular threshold, $t=1$, does not perform well~\cite{10.1145/3372297.3420013}. 
They recommend small thresholds $>$1, such as 2 to 15; we used a 10\% (5--7) threshold.
Others analysed a small sample of 9k apps from 9 third-party Android markets with a threshold of 6 and found 5\% apps were malicious~\cite{Buchanan2017}. 31\% had not been analysed by VirusTotal, yielding no results.
Furthermore, we found a higher proportion of malicious apps in modded markets, and compared modded apps to their Google Play counterparts.
Most approaches use a similar approach with different thresholds~\cite{blackbox_vt}. 
Others used weighted voting, relied on supervised learning, and used future results (after 4 weeks) as ground truth~\cite{malware_ground}. Others confirmed the increased accuracy of older results~\cite{maat,malwarelabeling}. 
%
%
%
Others focused on native code libraries misuse in a small sample of Android apps from one third-party market~\cite{wang2017nativespeaker}. It required manual verification for some types of misuse, unsuitable at scale. 
Similarly, others identified harmful libraries in Android and iOS based on VirusTotal results~\cite{chen2016crossplatformlibraries}. Our study links the presence and changes of ad libraries with changes in ad IDs and permissions.

Previous research has explored sideloading user motivations and knowledge~\cite{sideloading_motivations}, but they did not consider modded apps nor motivations of the maintainers. Their questionnaire is run on a sample of Computer Science students and staff, as well as relevant sideloading and rooting Reddit forums users, thus providing limited data on the real-world occurrence of sideloading.

\section{Conclusion}
This paper presented the results of the first large-scale study into modded app markets,  
with a large-scale technical analysis of 146k modded apps available on the 13 most popular markets. By comparing them to their Google Play counterparts, we demonstrated the vast majority were modified in one or more ways, including those labelled as unmodified. 
Our updated dataset with 300k apps is publicly available.
Modded markets likely reduce developers' and official markets' income due to pirated apps and unlocked premium features. 
The majority of apps fell into the gaming category, however many other popular apps exist on these markets, including a modified version of TikTok advertised as offering free coins and a modified version of Spotify offering ad-free music without subscription.
Modded apps add permissions and ad libraries; 21\% had different ad IDs than their Google Play version, suggesting ad revenue may be diverted away from original developers. 

From the users' perspective, modded apps advertise new, desirable features, which our case studies show often, although not always, work.
However, these markets are unrelated to the genuine developers, and divert or curtail the app purchase, IAPs and ads revenue streams.
While ad-free versions of apps are widely touted, fewer than $3\%$ of modded apps had all ads and trackers removed.
VirusTotal marked 9\% of code-modded apps as containing adware, grayware or Trojans, 10 times the rate found in Google Play. 
Modded markets continue to host malicious apps removed from Google Play for a long time. Users risk their personal data and their privacy being breached by downloading apps which code has not been verified by any official entity and modified by third-parties.
Furthermore, users might put others' privacy and security at risk, as modded apps might allow private content to be viewable by others, malware and spyware might make use of added permissions to access other users' private information, contacts, etc. 

Developers should be aware of these markets and practices, and 
given more tools and support to find and report malicious versions of their apps. 
Developers, especially smaller ones, will have a hard time reporting misuse of their intellectual property since at present they would need to manually report multiple versions of their apps in more than 400 modded Android markets. With current legislation such as DMCA, this only covers those versions flagged, so with every app update appearing in the markets they would have to repeat the process. 

The question of whether and how mobile devices should allow installing apps outside the official market is under investigation by regulators in the EU 
and the UK. 
Android offers official support, while iOS makes it very hard for the average consumer to install apps from outside the official market. 
Our work suggests regulators should consider options to counter the negative effects of modded markets and sideloading while protecting or enhancing user and app developer choice and protection. 
There are a range of options, including allowing sideloading while requiring apps to be tested and signed by an approved tester; requiring the distribution of alternative market apps through the official market in order to offer a pinch-point to support regulation; etc.
A confounding factor is that large revenue streams are tied to the status quo where a percentage of the price of paid apps and IAPs flow to the official market operator.

%

\section*{Acknowledgment}
We thank Richard Clayton for his help in setting up and maintaining the scraping machines, proxies and storage. We thank Stan (Jiexin) Zhang for his contributions to some scrapers. We are grateful to VirusTotal for the academic access to their API. We thank colleagues and anonymous reviewers for their feedback. This work is supported by Nokia Bell Labs (for LAS and ARB) and the European Research Council under the Horizon 2020 programme (grant agreement No 949127) (for HSD and AH). 


%
%
%
\bibliographystyle{IEEEtran}
\bibliography{references}

\end{document}